\documentclass[sigconf,nonacm]{acmart}

\AtBeginDocument{%
  }

\usepackage{amsmath}
\usepackage{amsfonts}

\usepackage{bbm}

\usepackage{tablefootnote}
\usepackage{diagbox}
\usepackage{booktabs}
\usepackage{multirow}

\usepackage{soul}
\usepackage{xcolor}
\newcommand{\ctext}[2]{%
  \begingroup
  \sethlcolor{#1}%
  \hl{#2}%
  \endgroup
}

\usepackage{threeparttable}

\usepackage{colortbl}

\begin{document}

\title[Guaranteeing Faithful Evidence Extraction in Speculative Retrieval-Augmented Generation]{Guaranteeing Faithful Evidence Extraction in Speculative Retrieval-Augmented Generation}

\author{Quentin Sign\'e}
\orcid{0009-0003-7569-1803}
\affiliation{%
  \institution{Université de Toulouse - IRIT UMR 5505}
  \city{Toulouse}
  \country{France}
}
\affiliation{%
  \institution{Airbus Protect}
  \city{Blagnac}
  \country{France}
}
\email{quentin.signe@airbus.com}

\author{Mohand Boughanem}
\orcid{0000-0001-7004-0807}
\affiliation{%
  \institution{Université de Toulouse - IRIT UMR 5505}
  \city{Toulouse}
  \country{France}
}
\email{mohand.boughanem@irit.fr}

\author{Jose G. Moreno}
\orcid{0000-0002-8852-5797}
\affiliation{%
  \institution{Université de Toulouse - IRIT UMR 5505 }
  \city{Toulouse}
  \country{France}
}
\email{jose.moreno@irit.fr}

\author{Thiziri Belkacem}
\orcid{0000-0001-9454-0996}
\affiliation{%
  \institution{Airbus Protect}
  \city{Blagnac}
  \country{France}
}
\email{thiziri.belkacem@airbus.com}

\renewcommand{\shortauthors}{Signé et al.}

\begin{abstract}
Large Language Models (LLMs) are increasingly used as interfaces for information retrieval, but they remain prone to hallucinations and faithfulness errors, in which the generated answers diverge from the retrieved evidence. While Retrieval-Augmented Generation (RAG) and recent hybrid or semi-extractive approaches mitigate this issue, they do not guarantee that quoted or extracted spans are verbatim from the retrieved context. Furthermore, current speculative decoding methods, even when extracting spans from retrieved documents, primarily prioritise inference efficiency over faithfulness. This limitation can have severe consequences in safety-critical domains, where answers must exactly match certified documentation.

We introduce Constrained Hybrid Decoding (CHyD), a novel faithfulness-first paradigm for speculative RAG. While traditional speculative decoding is optimised for inference speed, CHyD repurposes this architecture to ensure faithful verbatim evidence extraction when the extraction mode is correctly triggered. Our approach enforces hard decoding constraints that restrict generation to continuous spans present in the retrieved documents. This design provides a robust but straightforward guarantee:  any explicitly quoted span in the output appears verbatim in the provided context. Unlike semi-extractive QA (e.g., SEMQA) and speculative RAG (e.g., NEST), our approach prioritises extraction correctness as its primary objective.

We evaluate our method across state-of-the-art LLMs on diverse abstractive, extractive, and semi-extractive QA benchmarks, including technical datasets motivated by aircraft maintenance. Results show that existing hybrid methods frequently hallucinate quoted spans, with exact extraction accuracy dropping below $40\%$ in technical domains. In contrast, our approach achieves near-perfect extraction faithfulness regardless of the model used. Although enforcing hard constraints introduces a trade-off with fluency-oriented metrics, our method improves exact answer correctness and remains competitive overall, highlighting its suitability for safety-critical information retrieval applications.
\end{abstract}


\maketitle

\section{Introduction}
Large Language Models (LLMs) have become a central component of modern Question-Answering (QA) systems. Despite their impressive generative capabilities, LLMs are prone to generating ungrounded content that appears factual \cite{tonmoy2024comprehensive}, a phenomenon known as hallucination \cite{ji2023survey}. In document-grounded QA, this manifests as context inconsistencies, where the generated answer contradicts or deviates from the retrieved documents \cite{huang2025survey}. While such errors may be tolerable in open-domain applications, they are unacceptable in safety-critical fields, such as medicine, law, and aircraft maintenance. In these industrial and regulated settings, Question-Answering systems are expected not only to provide plausible answers but also to return exact procedures, thresholds, or conditions from certified documentation. For instance, in aircraft maintenance, operations must strictly follow approved manuals, and paraphrasing or abstracting may invalidate procedural compliance.
Indeed, in aircraft maintenance operations, any use of AI systems and solutions must demonstrate specific safety and compliance guarantees. Hence, QA systems for decision-making support should rigorously verify their answers. These answers must not only be correct but also verbatim and faithful to the retrieved context. \citet{huang2025survey} define this concept of faithfulness as the absence of divergence between the generated content and the supporting context.

Traditional Retrieval-Augmented Generation (RAG) approaches mitigate the risk of hallucination by grounding answers in external documentation \cite{gao2023retrieval, shuster2021retrieval}, thus improving factuality. However, standard RAG frameworks still rely on probabilistic token generation and therefore cannot fully guarantee that the generated answers are faithful to the retrieved context. Recent work has explored hybrid paradigms that combine generation with extraction, including semi-extractive QA \cite{schuster2024semqa} and speculative RAG methods \cite{zhao2026retrieval}. By design, these approaches provide answers that are partially generated and partially quoted from the context, improving factuality on average, but without guaranteeing verbatim extraction.

This paper addresses a fundamental gap in the literature. To our knowledge, no hybrid generation framework is primarily designed to guarantee faithfulness rather than fluency or inference speed. Existing speculative RAG frameworks (e.g. NEST \cite{li2024nearest}, REST \cite{he2024rest}) use copying mechanisms to accelerate decoding \cite{zhao2026retrieval}. Semi-extractive methods such as SEMQA \cite{schuster2024semqa} rely on the model to generate quoted spans, which can still hallucinate. None of these approaches ensures that the generated evidence is extracted verbatim from the context.

In safety-critical industries like aircraft maintenance, the research challenge is not only to generate a \textit{``correct''} answer but also to ensure verifiable compliance with certified documentation. We propose Constrained Hybrid Decoding (CHyD), a faithfulness-first paradigm for speculative RAG frameworks that addresses this challenge by enforcing robust safety constraints during decoding. When the model enters extraction mode, decoding is restricted to tokens that form a continuous, verbatim span from the retrieved documents. This design provides a robust yet straightforward guarantee: any quoted span in the output is exactly extracted from the provided context. Unlike prior speculative RAG approaches that use copying for efficiency, CHyD uses constrained decoding to enforce faithful evidence extraction. We demonstrate that this guarantee leads to near-perfect extraction correctness across diverse QA benchmarks, at the cost of modest reductions in fluency-oriented metrics.

Our contributions are:
\begin{itemize}
    \item Constrained Hybrid Decoding (CHyD), a speculative RAG framework designed to provide faithfulness guarantees;
    \item An extensive evaluation demonstrating that our approach outperforms prior speculative and semi-extractive methods in extractive faithfulness, particularly in technical and safety-critical domains;
    \item SEMAeroSQuAD, a dataset tailored for semi-extractive QA evaluation in safety-critical contexts.
\end{itemize}

\begin{figure*}
    \centering
    \includegraphics[width=\linewidth]{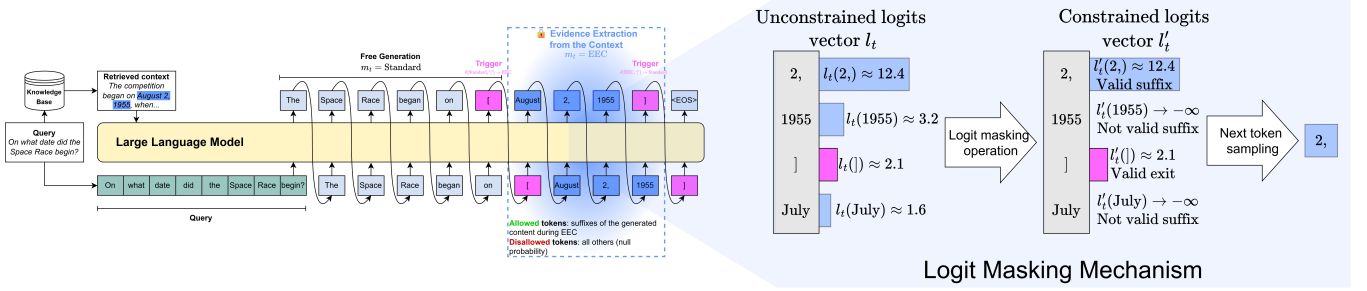}
    \Description{The figure illustrates our approach, Constrained Hybrid Decoding (CHyD), a hybrid decoding framework in which a language model alternates between free-text generation and the constrained extraction mode. This extractive mode allows only verbatim spans copied from the retrieved context.}
    \caption{Constrained Hybrid Decoding (CHyD) for faithful evidence extraction. In Standard mode ($m_t=\text{Standard}$), the model generates text in a free and autoregressive manner. The transition function $\delta$ switches the model into extraction mode ($m_t=\text{EEC}$) when the trigger token \texttt{``[''} is generated. During EEC mode, the model produces a constrained logits vector $l'_t$, which preserves the original probabilities $l_t$ for valid continuous verbatim spans (e.g. ``2,'') and the closing trigger (\texttt{``]''}), while assigning $-\infty$ (null probability) to all disallowed tokens (e.g. ``1955'' and ``July''). Free generation resumes when $\delta$ is triggered by \texttt{``]''}.}
    \label{fig:diagram}
\end{figure*}

\section{Related Work}

A critical challenge in Large Language Models (LLMs) is faithfulness hallucination, in which the generated outputs diverge from the supporting evidence. In document-grounded Question-Answering (QA) tasks, this issue manifests as context inconsistencies, where the model's answer contradicts or deviates from the retrieved evidence \cite{huang2025survey}. In safety-critical domains, even minor deviations from the certified documentation are unacceptable, motivating approaches that strictly align outputs with the document's content to ensure procedural compliance. 

To mitigate these risks, Retrieval-Augmented Generation (RAG) \cite{lewis2020retrieval} has become the standard approach. This paradigm enables LLMs to ground their answers in external documents, thereby improving accuracy and faithfulness. While recent RAG approaches employ optimisations such as improved retrieval steps \cite{yao2023react, shi2024replug} or fine-tuning \cite{borgeaud2022improving, sharma2024retrieval}, they do not fully mitigate hallucinations \cite{tonmoy2024comprehensive}. Because these methods rely on unconstrained probabilistic token generation, they remain prone to inconsistencies and cannot guarantee that the generated answers faithfully reproduce the retrieved evidence. Consequently, hallucinations remain a significant challenge, especially in technical domains where language model training data lacks specialised vocabulary \cite{sharma2024retrieval}.

Recent research has attempted to improve faithfulness by directly manipulating the decoding process. Logit-based RAG approaches dynamically adjust the model's output token distribution by weighting parametric knowledge against contextualised knowledge \cite{zhao2026retrieval}. For instance, kNN-LM \cite{khandelwal2020generalization} combines the probability distributions of an LLM and a kNN model, and subsequent approaches like Context-aware Decoding (CAD) \cite{shi2024trusting}, AdaCAD \cite{wang2025adacad}, and CoCoA \cite{khandelwal2025cocoa} aim to reduce unfaithful generation by dynamically combining the model's prior and contextualised knowledge. Similarly, Pointer-Generator Networks \cite{see2017get} introduce a soft copying mechanism that allows the model to use tokens from the source document during generation. While these approaches reduce context inconsistencies and unfaithful generation on average, they enforce soft faithfulness preferences rather than guaranteeing faithfulness. Because these methods rely on probabilistic weighting rather than hard constraints, the generated answer may still contain spans that appear grounded but are not extracted verbatim from the retrieved documentation.

Conversely, recent research has also studied purely extractive approaches that restrict the output to spans from retrieved documents, including QA models built on encoder-only architectures such as BERT \cite{devlin2019bert}. The SEBRAG \cite{signe2025substring} framework utilises a RAG pipeline in which the LLM integrates a substring extraction tool, while \citet{mallick2023adapting} adopt a span-prediction method which predicts correct token indices by prompting a language model. Although these methods aim to reduce context inconsistencies by providing strong extraction guarantees, they enforce verbatim copying from the retrieved context. As a result, they lack the fluency and synthesis capabilities required for complex tasks like multi-source QA.

Extractive and abstractive methods have been studied separately \cite{luo2022choose}, but to balance fluency and grounding, hybrid frameworks have been developed to produce answers that are partially generated and partially extracted. \citet{cheng2021unitedqa} propose a hybrid approach that combines the outputs of extractive and abstractive readers, achieving strong performance compared to either reader alone. More recently, SEMQA \cite{schuster2024semqa} introduces a semi-extractive format that combines free text with quoted spans, along with the open-domain QuoteSum dataset and an associated evaluation metric designed to assess hybrid answers. However, these ``quoted-generation'' approaches \cite{worledge2024extractive} rely on the LLM to generate the quoted span (e.g. between brackets \cite{schuster2024semqa}) rather than on pure evidence extraction. In these frameworks, because quoted spans are freely generated rather than copied, the model can mimic citation formats while still hallucinating the quoted content. Therefore, generated answers may be fluent, while the quoted spans they contain are absent from the retrieved context. Consequently, SEMQA \cite{schuster2024semqa} optimises answer quality under a semi-extractive format, but does not enforce the copying of continuous spans from the retrieved context, hence failing to prevent context inconsistencies.

Other related works enforce constraints directly during decoding to control the content of the generated output. Lexically Constrained Decoding methods, such as Grid Beam Search \cite{hokamp2017lexically}, ensure that predefined lexical constraints appear in the answer by modifying the beam search procedure to track constraint satisfaction. These approaches are effective at enforcing the inclusion of specific tokens, but they do not guarantee that the generated spans are continuous substrings of the retrieved documents.

To bridge the gap between extraction and generation, speculative RAG \cite{zhao2026retrieval} dynamically alternates between generating tokens freely and copying spans from retrieved external content. Speculative decoding traditionally relies on a draft model to propose candidate tokens, which are then verified by a larger model \cite{leviathan2023fast}. Speculative RAG adapts this paradigm by using retrieved documents to propose draft tokens instead of a separate draft model. While quoted-generation methods rely on the model to generate evidence in a constrained format, copy-based speculative RAG approaches directly reuse verbatim text from the context during generation. Methods like CoG \cite{lan2023copy}, REST \cite{he2024rest}, and CopySpec \cite{dumitru2025copyspec} use these ``copy-paste'' operations during decoding, producing hybrid outputs that are partially generated and partially extracted from the context, thereby improving factuality, faithfulness, and transparency. 
Similarly, NEST \cite{li2024nearest} combines a logit-based and speculative decoding approach to copy retrieved spans at judicious decoding steps. However, the primary objective of these frameworks is to accelerate inference by bypassing the traditional autoregressive token-by-token generation \cite{somasundaram2025pld}. Indeed, copying a span from the context rather than generating tokens individually can drastically reduce end-to-end generation latency. This ``copy-paste'' mechanism allows these approaches to accelerate inference, but without guaranteeing correctness \cite{li2024nearest}. 
Although effective for their intended purpose, these existing methods are not designed to meet the strict requirements of safety-critical domains, such as aircraft maintenance, where it is necessary to prioritise accuracy and traceability over latency to ensure that the correct procedures are followed. The key distinction between CHyD and prior speculative RAG approaches (particularly NEST) lies not in the use of copying itself, but in the role that copying plays during decoding. Existing speculative approaches employ copying as a latency optimisation strategy, whereas CHyD uses constrained copying as a correctness guarantee mechanism.

In summary, existing approaches either improve faithfulness through soft probabilistic mechanisms without strong guarantees, enforce faithfulness through purely extractive strategies at the expense of fluency, or employ hybrid or speculative decoding mechanisms that prioritise efficiency over evidence correctness.

To the best of our knowledge, no prior work explicitly enforces verbatim evidence extraction as a decoding constraint within a hybrid QA framework. This gap motivates our approach, CHyD, which reinterprets the speculative RAG paradigm to prioritise faithfulness over inference speed. Indeed, rather than using speculative mechanisms solely for efficiency, CHyD transforms speculative decoding into a constrained evidence-verification process, leveraging dynamic switching and hard lexical constraints to ensure that extracted evidence corresponds to valid continuous verbatim spans from the retrieved context. This design enables trustworthy hybrid QA systems, particularly suitable for safety-critical and compliance-sensitive environments.

\begin{table*}[ht]
    \caption{Overview of the evaluation datasets. Lengths were computed using the Qwen3 tokeniser.}
    \label{tab:datasets}
    \centering
    \resizebox{\textwidth}{!}{%
    \begin{tabular}{l cccc}
    \toprule
    \bf Characteristic & \bf MedMCQA & \bf MESAQA & \bf QuoteSum & \bf SEMAeroSQuAD \\
    \midrule
     Source & Sample of MedMCQA \cite{pal2022medmcqa} & Sample of MESAQA \cite{wang2025mesaqa} & QuoteSum \cite{schuster2024semqa} & Sample of SQuAD 2.0 \cite{rajpurkar2018know} \\
     Domain & Medical & Medical & Open-domain & Aeronautics and Space \\
     Task Type & Extractive QA & Abstractive QA & Semi-extractive QA & Semi-extractive QA\\
     Number of Questions & 2816 & 2999 & 451 & 1420 \\
     Avg. Context Length (tokens) & 4.9 & 831.4 & 138.9 & 166.3 \\
     Avg. Question Length (tokens) & 19.5 & 11.4 & 10.4 & 13.2\\
     Avg. Answer Length (tokens) & 5.1 & 62.9 & 64.3 & 16.7 \\
     \% Unanswerable Questions & 0.0 \% & 0.0 \% & 0.0 \% & 24.6 \% \\
    \bottomrule
    \end{tabular}
    }    
\end{table*}

\section{Approach}

While soft faithfulness objectives may be sufficient for open-domain QA, safety-critical applications require even stronger guarantees of faithfulness. We therefore propose a framework that enforces extraction correctness at decoding time, rather than relying on post-hoc validation or soft preferences.

Our proposed constrained hybrid generation approach (Figure \ref{fig:diagram}), designed for safety-critical domains where faithfulness guarantees are necessary, leverages LLM strengths by introducing a switching mechanism that enables the model to generate free-form text and extract parts of the retrieved documentation.

Given an input query $q$, a pre-trained Language Model (LM) $\mathcal{M}$, a corpus $\mathcal{D}$, and the $k$ retrieved contexts $\mathcal{C}=\{c_1,c_2,\dots,c_k\}$ from $\mathcal{D}$ with respect to $q$, we define the decoding process as follows:

In a standard Retrieval-Augmented Language Model setup, at each generation step $t$, the LM $\mathcal{M}$ produces a vector of logits $l_t \in \mathbb{R}^{|V|}$, where $|V|$ is the size of the model's vocabulary. The next token $y_t$ is then sampled from the probability distribution given by the softmax of $l_t$:
\[
    p_{\mathcal{M}}(y_t|q,\mathcal{C},y_{<t}) = \text{softmax}(l_t)
\]
To enable hybrid generation, we introduce two decoding modes in which the LM can be at any step $t$:
\begin{enumerate}
    \item \textbf{Standard Mode}: The LM generates text using the standard decoding process as described above.
    \item \textbf{Extraction of Evidence from Context (EEC) Mode}: The LM is forced to generate a sequence of tokens that forms a continuous verbatim span within at least one passage $c_i \in \mathcal{C}$.
\end{enumerate}

When the model enters extraction mode, decoding is restricted to tokens that form a continuous, verbatim span within the retrieved documents, ensuring the expected level of faithfulness in safety-critical contexts.
To formalise the switching mechanism, we model decoding as a two-state process. We define $m_t \in \{\text{Standard}, \text{EEC}\}$ as the model's decoding mode at step $t$.
The mode is updated at each step via a transition function $m_t = \delta(m_{t-1},y_{t-1})$ defined as:
\[
\delta(m,y) = \left\{
    \begin{array}{ll}
        \text{EEC} & \text{if}~m = \text{Standard}~\text{and}~y=\texttt{``[''} \\
        \text{Standard} & \text{if}~m = \text{EEC}~\text{and}~y=\texttt{``]''} \\
        m & \text{otherwise}
    \end{array}
\right.
\]

The transition between the two generation modes is managed by two specific tokens: the model enters the extraction process when \texttt{``[''} is produced and exits it when \texttt{``]''} is triggered. This design choice enables a simple, model-agnostic switching mechanism that, once extraction mode is entered, enforces faithfulness during decoding via hard constraints. Moreover, this design allows future work to explore alternative trigger conditions. 

Generation terminates when the model produces the \textit{end-of-sequence} (EOS) token, while in Standard mode ($m_t=\text{Standard}$), or when a predefined maximum number of new tokens is reached. EOS token generation is not allowed during EEC mode to ensure quoted spans are closed before generation completes.

To enforce the verbatim constraint, we first define a token selection function $f$. Let $S=\{v_1, v_2, \dots, v_n\}$ be a token sequence generated by the model after entering the EEC mode. The function $f(S,\mathcal{C})$ returns the set $\mathcal{F}$ of all valid next tokens from the vocabulary $V$ such that the new extended sequence is an exact verbatim match in at least one passage $c_i \in \mathcal{C}$.
\begin{equation*}
\begin{split}
    \mathcal{F} = f(S,\mathcal{C})=\{v' \in V \mid \exists c_i \in \mathcal{C}~\text{such that the sequence} \\
    \{v_1, v_2, \dots, v_n,v'\} ~\text{exists as a continuous span of}~c_i \}
\end{split}
\end{equation*}

In rare cases, the set of valid next tokens $\mathcal{F}$ may be empty, for instance, when the end of the context is reached during EEC mode, and consequently, no continuous token matches the generated prefix. In these situations, we exit extraction mode by generating the special token \texttt{``]''} and resume standard generation mode.

When the model is in EEC mode at step $t$ (i.e. $m_t=\text{EEC}$), having entered this mode at step $t_{open}$, the generation proceeds as follows:

\begin{enumerate}
    \item The sequence already generated within the EEC mode is defined as $S_t=\{y_{t_{open}}, y_{t_{open}+1}, ..., y_{t-1} \}$.
    \item The set of next possible tokens $\mathcal{F}_t = f(S_t, \mathcal{C})$ is retrieved.
    \item The LM $\mathcal{M}$ produces its standard logits vector $l_t$.
    \item The constraint mask is applied to create a new logits vector $l'_t$ defined for every token $v \in V$ as:
        \[
        l'_t(v) = \left\{
            \begin{array}{ll}
                l_t(v) & \text{if}~v \in \mathcal{F}_t~\text{or}~v=\texttt{``]''}   \\
                -\infty & \text{otherwise}
            \end{array}
        \right.
        \]

    This masking step ensures that the model can either continue extracting a verbatim span from the context or exit EEC mode and resume free generation by producing the special token \texttt{``]''}.
    Moreover, this step allows us to set the probability of sampling any token not in $\mathcal{F}_t$ to zero.
    \item Finally, the next token $y_t$ is sampled from the new probability distribution $p_{\mathcal{M}}(y_t|q,\mathcal{C},y_{<t}) = \text{softmax}(l'_t)$ derived from the constrained logits vector $l'_t$.
\end{enumerate}

The CHyD mechanism ensures that, in the EEC mode, generated tokens either match a verbatim span from the retrieved context or explicitly terminate the extraction process. 
Moreover, this mechanism allows the model to switch between the two modes multiple times, enabling several extractions within the same answer across different passages of the retrieved context. This strong constraint distinguishes our approach from prior hybrid methods such as SEMQA \cite{schuster2024semqa} and NEST \cite{li2024nearest}, which rely on soft constraints or probabilistic span acceptance. In contrast, CHyD enforces verbatim evidence extraction during decoding, making it particularly suited to safety-critical QA tasks.

\section{Experimental Setup}

We evaluate our faithfulness-first speculative RAG framework, CHyD, across multiple Question-Answering tasks to assess its ability to generate fluent answers while ensuring precise extraction of supporting evidence. All methods are evaluated under a controlled and consistent experimental setup to ensure fair comparisons.

\begin{table*}[ht]
    \caption{Examples of question-answer pairs from the evaluation datasets. Semi-extractive answers contain quoted spans (highlighted in \ctext{blue!30}{[ blue ]}) extracted from the retrieved context. The associated contexts are not displayed for clarity.} 
    \label{tab:qa_examples}
    \centering
    \resizebox{\textwidth}{!}{%
    \begin{tabular}{lp{0.3\textwidth}p{0.55\textwidth}}
    \toprule
    \bf Dataset & \bf Question & \bf Answer \\
    \midrule
    MedMCQA & \textit{Characteristic of venous blood flow of lower limb in duplex Doppler is?} & monophasic \\
    \addlinespace
     MESAQA & \textit{How does physical therapy and exercise help with rheumatoid arthritis?} & Physical therapy and exercise can help people with rheumatoid arthritis to move better and with less pain. Moist heat, ice packs, and relaxation techniques can also help ease symptoms. Occupational therapy can help people learn how to do everyday activities. \\
     \addlinespace
     QuoteSum & \textit{When does the world cup start and end?} & Some sources state that the \ctext{blue!30}{[ FEI World Cup Jumping 2011/2012 ]} \ctext{blue!30}{[ tournament series ]} ran \ctext{blue!30}{[ from October 12, 2011 to February 26, 2012. ]} Others state that \ctext{blue!30}{[ The 2017-18 UCI Track Cycling World Cup ]} \ctext{blue!30}{[ series was run from 3 November 2017 to 21 January 2018 ]} , that \ctext{blue!30}{[ The 2016 FINA Swimming World Cup ]} \ctext{blue!30}{[ started 13 days after the final day of the Olympic pool swimming program ]} , and that \ctext{blue!30}{[ 2017 Canoe Slalom World Cup ]} \ctext{blue!30}{[ opened ]} \ctext{blue!30}{[ June 16-18 ]} \ctext{blue!30}{[ and concluded ]} in \ctext{blue!30}{[ September 8-10 ]} . \\
     \addlinespace
     SEMAeroSQuAD & \textit{Radar was supplemented by what in the 1980s?} & In the 1980s, radar was supplemented by \ctext{blue!30}{[ optronics ]} .\\
    \bottomrule
    \end{tabular}
    }
\end{table*}

\subsection{Datasets}

We evaluate CHyD across multiple Question-Answering datasets, covering fully abstractive, extractive and semi-extractive settings. This diversity enables the analysis of trade-offs between fluency, answer quality and faithfulness. 

In addition, we introduce SEMAeroSQuAD, a semi-extractive dataset designed to evaluate faithfulness-critical QA in technical and safety-sensitive domains.
Each dataset is categorised by the nature of its answers: fully generated answers (abstractive), verbatim spans (extractive), or hybrid answers that combine free text with explicitly marked evidence (semi-extractive). Table \ref{tab:datasets} summarises the datasets' characteristics, and Table \ref{tab:qa_examples} provides some examples.

\textbf{Abstractive QA.} We use MESAQA \cite{wang2025mesaqa}, a medical domain-specific dataset that requires free-form abstractive answers grounded in clinical documents. This dataset assesses the global quality of answers in terms of fluency when utilising faithfulness constraints.

\textbf{Extractive QA.}  We evaluate extractive grounding on MedMCQA \cite{pal2022medmcqa}, a multiple-choice medical QA dataset. Because MedMCQA gold answers are verbatim spans from the context, we prompt models to answer in complete sentences that incorporate the selected option. We use this dataset exclusively to assess extractive accuracy, enabling us to evaluate the faithfulness of the quoted spans. MedMCQA is also used to evaluate the extractive correctness of speculative RAG methods such as NEST \cite{li2024nearest}, making it a relevant benchmark for comparison.

\textbf{Semi-extractive QA.} We also use QuoteSum \cite{schuster2024semqa}, an open-domain dataset in which answers include extracted spans from several passages, denoted by brackets. QuoteSum evaluates hybrid generation, including the correctness of the extracted spans, and the overall quality of the answers. QuoteSum is one of the contributions of \citet{schuster2024semqa}, who introduced the SEMQA framework, enabling direct comparison with prior semi-extractive approaches.

\textbf{SEMAeroSQuAD: A semi-extractive closed-domain dataset.} Existing technical QA benchmarks do not assess whether quoted spans are extracted verbatim from the retrieved documents, a mandatory requirement in safety-critical domains such as aircraft maintenance. To address this gap, we introduce SEMAeroSQuAD\footnote{\url{https://github.com/quentin-sgn/Constrained-Hybrid-Decoding}}, a semi-extractive dataset derived from SQuAD 2.0 \cite{rajpurkar2018know} and focused on aeronautical and space-related content.
SEMAeroSQuAD is not intended as a new benchmark, but rather as a task-oriented test set designed to evaluate faithfulness in hybrid generation within closed-domain settings. Its primary objective is to assess whether models can reproduce evidence verbatim from the context while generating fluent surrounding content, reflecting the requirements of safety-critical applications. 
To construct SEMAeroSQuAD, we follow the construction process of AeroSQuAD \cite{signe2025substring} by selecting aeronautical and space-related questions from SQuAD 2.0. We then transform their extractive answers into semi-extractive responses. The original extracted gold answer is preserved and embedded within a fluent and grammatically correct sentence, while the surrounding text is generated using Gemini 2.5 \cite{comanici2025gemini}. This procedure ensures that each example contains the original gold answer from SQuAD 2.0, which must be reproduced in the answer to be considered faithful. For instance, for the query \textit{``On what date did the Space Race begin?''}, the original answer in the SQuAD dataset is \textit{``August 2, 1955''}, which is embedded in our dataset, SEMAeroSQuAD, as \textit{``The Space Race began on [ August 2, 1955 ]  when the Soviet Union announced it would also launch a satellite.''}. Hence, the extracted answer is integrated into a complete, grammatically correct answer and is denoted by the brackets surrounding it. We acknowledge that using an LLM to generate surrounding text introduces biases. However, as the quoted spans remain unchanged from SQuAD 2.0, SEMAeroSQuAD provides a controlled setting to evaluate extraction correctness in closed-domain QA.

\subsection{Baselines}

We compare our faithfulness-focused framework, CHyD, against the closest prior methods for hybrid generation and speculative RAG. SEMQA \cite{schuster2024semqa} is a semi-extractive QA approach that combines free-form text with generated quoted spans, optimising answer quality without enforcing extraction correctness. NEST \cite{li2024nearest} is a speculative RAG method that pastes retrieved spans to accelerate inference by combining a logit-based approach with relaxed speculative decoding. To establish a purely extractive upper bound for our evaluation metrics, we also include the encoder-only DistilBERT-SQuAD \cite{sanh2019distilbert} baseline, fine-tuned on SQuAD v1.1. Because this model predicts start and end token indices rather than generating text probabilistically, it provides a reference for extraction accuracy.

\subsection{Evaluation Metrics}

We evaluate the approaches against several criteria to assess answer correctness, generation quality, fluency, and faithfulness to the retrieved evidence. This combination of evaluation criteria will help us highlight potential trade-offs in hybrid QA and enable more precise analysis, particularly in safety-critical settings.

To evaluate the methods' evidence extraction performance, we report Exact Match (EM) and F1 scores for quoted spans. These are standard measures of answer correctness, allowing us to assess both absolute precision (EM) and partial overlap (F1) with the gold reference. Following prior work on speculative RAG \cite{li2024nearest}, we also compute Answer-Level Recall, which checks whether the extracted parts of the output contain any correct verbatim answers. 

To evaluate fluency and answer quality, we report ROUGE-L \cite{lin2004rouge}, BERTScore \cite{zhang2019bertscore}, and MAUVE \cite{pillutla2021mauve}. This combination provides a global assessment of answer quality. ROUGE-L measures lexical overlap and sentence structure, BERTScore uses contextual embeddings to assess semantic similarity, and MAUVE evaluates the overall distribution of the generated text to estimate how closely it resembles human-written answers. We also compute the SEMQA score \cite{schuster2024semqa}, which combines generation quality (ROUGE-L) and extraction performance (F1 score of the answer's quoted spans). These metrics capture semantic similarity and generation quality but do not assess whether the quoted spans are verbatim from the retrieved context. Moreover, generation metrics such as ROUGE or BERTScore reward fluent, semantically plausible answers, even when evidence is generated rather than copied. As a result, models may achieve high scores while producing unfaithful answers.

To directly evaluate one of the core objectives of safety-critical QA, we introduce Extraction Faithfulness Accuracy (EFA). EFA is a precision-oriented metric for the EEC mode that measures the proportion of quoted spans in the generated answer that appear verbatim in the retrieved context.
Formally, let $\mathcal{S}=\{s_i\}_{i=1}^{N}$ denote the set of all quoted spans produced across the evaluation set. Each span $s_i$ is associated with a set of contexts $\mathcal{C}_i = \{c_{i,1}, \dots, c_{i,k}\}$ retrieved for the associated query. A quoted span $s_i$ is counted as faithful if it appears as a continuous substring of at least one of the retrieved contexts. Then, EFA is defined as:
\[
    \mathrm{EFA} = \frac{1}{N} \sum_{i=1}^{N} \mathbbm{1}[\exists c \in \mathcal{C}_i \mid s_i \subseteq c]
\]
where $\mathbbm{1}[\cdot]$ is an indicator function equal to 1 if the quoted span $s_i$ appears as a continuous span in at least one $c \in \mathcal{C}_i$, and 0 otherwise. By construction, $\text{EFA} \in [0,1]$, with higher values indicating stronger guarantees that the quoted span is extracted from the retrieved context.

Unlike EM and F1, EFA directly assesses whether quoted evidence is truly extracted rather than merely copied. Consequently, EFA penalises partially correct or paraphrased quotations and naturally accounts for answers containing multiple quoted spans. 

To ensure rigorous evaluation, we use the McNemar test for binary metrics (EM and ALR) and the Wilcoxon signed-rank test for continuous metrics, and apply the Holm-Bonferroni correction to all raw p-values.

\begin{table}[ht]
    \caption{Generation parameters used by the models for each dataset. All methods used the same configuration per dataset, and the generation parameters were chosen based on state-of-the-art standards and dataset-specific characteristics.}
    \label{tab:generation_config}
    \centering
    \resizebox{\columnwidth}{!}{%
    \begin{tabular}{l cccc}
        \toprule
        \bf Dataset & \bf Temp. & \bf Top-p & \bf Max Tokens & \bf Few-shots \\
        \midrule
        MedMCQA    & 0.7 & 0.9 & 64  & 1 \\
        MESAQA     & 0.7 & 0.9 & 128 & 1 \\
        QuoteSum   & 0.7 & 0.9 & 256 & 2 \\
        SEMAeroSQuAD  & 0.7 & 0.9 & 64  & 1 \\       
        \bottomrule
    \end{tabular}
    }
\end{table}

\begin{table*}[ht]
\caption{Extraction correctness and faithfulness across the datasets. Bold numbers indicate the best performance. EM: Exact Match. ALR: Answer Level Recall. EFA: Extraction Faithfulness Accuracy. $\dagger$ and $\ast$ denote significant improvement of CHyD over SEMQA and NEST, respectively ($p < 0.05$). Note: Llama-3.1 refers to Llama-3.1-8B-Instruct, Mistral-N refers to Mistral-Nemo-Instruct-2407, and Qwen3 refers to Qwen3-4B-Instruct.}
\label{tab:extractive_eval}
\centering
\resizebox{\textwidth}{!}{%
\begin{tabular}{ll | rrrr | rrcr | rrrr | rrrr | rrrr | rrrr}
\toprule
\bf LLM & \bf Approach & \bf EM & \bf F1 & \bf ALR & \bf EFA & \bf EM & \bf F1 & \bf ALR & \bf EFA & \bf EM & \bf F1 & \bf ALR & \bf EFA & \bf EM & \bf F1 & \bf ALR & \bf EFA & \bf EM & \bf F1 & \bf ALR & \bf EFA & \bf EM & \bf F1 & \bf ALR & \bf EFA \\
\midrule
\rowcolor{gray!10} & & \multicolumn{4}{c|}{\textbf{MedMCQA}} & \multicolumn{4}{c|}{\textbf{MESAQA}} & \multicolumn{4}{c|}{\textbf{QuoteSum}} &  \multicolumn{8}{c|}{\textbf{SEMAeroSQuAD}} & \multicolumn{4}{c}{\textbf{Overall}} \\
\rowcolor{gray!10} & & \multicolumn{4}{c|}{} & \multicolumn{4}{c|}{} & \multicolumn{4}{c|}{} &  \multicolumn{4}{c|}{single-context} &  \multicolumn{4}{c|}{multi-context} & \multicolumn{4}{c}{} \\
\addlinespace[0.5em]
DistilBERT  &   &  0.022  &  0.358  &  0.631  &  0.996  &   0.012  &  0.195  &  -  &  0.988  &  0.559  &  0.686  &  0.228  &  0.991  &  0.662  &  0.841  &  0.733  &  0.999  &  0.542  &  0.706  &  0.613  &  0.999 & 0.314 & 0.520 & 0.531 & 0.994 \\

\addlinespace[0.5em]
Llama-3.1  &  NEST  &  0.087  &  0.272  &  \textbf{0.524}  &  0.877  &  0.004  &  0.270  &  -  &  0.997  &  0.026  &  0.174  &  0.506  &  0.985  &  0.044  &  0.239  &  0.858  &  0.995  &  0.016  &  0.172  &  \textbf{0.893}  &  0.992  & 0.040 & 0.239 & 0.487 & 0.964 \\
  &  SEMQA  &  0.334  &  0.434  &  \textbf{0.524}  &  0.736  &  \textbf{0.037}  &  \textbf{0.362}  &  -  &  0.795  &  0.255  &  \textbf{0.434}  &  \textbf{0.643}  &  0.867  &  0.177  &  0.268  &  \textbf{0.868}  &  0.962  &  0.148  &  0.236  &  0.859  &  0.933  & 0.201 & \bf 0.374 & \bf 0.515 & 0.840 \\
  &  CHyD (ours)  &  \textbf{0.379}\rlap{$^{\ast\dagger}$}  &  \textbf{0.481}\rlap{$^{\ast\dagger}$}  &  0.491  &  \textbf{0.999}  &  0.028  &  0.301  &  -  &  \textbf{1.000}  &  \textbf{0.257}\rlap{$^{\ast}$}  &  0.415  &  0.621  &  \textbf{1.000}  &  \textbf{0.194}\rlap{$^{\ast}$}  &  \textbf{0.277}  &  0.865  &  \textbf{1.000}  &  \textbf{0.158}\rlap{$^{\ast}$}  &  \textbf{0.247}  &  0.855  &  \textbf{0.999}  & \bf 0.214 & 0.368 & 0.501 & \bf 1.000 \\
\addlinespace[0.5em]
Mistral-N  &  NEST  &  0.136  &  0.297  &  \textbf{0.489}  &  0.721  &  0.027  &  0.294  &  -  &  0.972  &  0.047  &  0.198  &  0.337  &  0.977  &  0.170  &  0.368  &  0.829  &  \textbf{0.983}  &  0.099  &  0.268  &  \textbf{0.818}  &  0.976  & 0.095 & 0.289 & \bf 0.446 & 0.913 \\
  &  SEMQA  &  0.309  &  0.424  &  0.395  &  0.628  &  0.026  &  \textbf{0.439}  &  - &  0.371  &  0.262  &  \textbf{0.421}  &  \textbf{0.428}  &  0.793  &  \textbf{0.369}  &  \textbf{0.618}  &  \textbf{0.833}  &  0.763  &  \textbf{0.299}  &  \textbf{0.516}  &  0.814  &  0.747  & 0.241 & \bf 0.475 & 0.426 & 0.639 \\
  &  CHyD (ours)  &  \textbf{0.415}\rlap{$^{\ast\dagger}$}  &  \textbf{0.529}\rlap{$^{\ast\dagger}$}  &  0.438  &  \textbf{0.981}  &  \textbf{0.063}\rlap{$^{\ast\dagger}$}  &  0.410  & -  &  \textbf{0.992}  &  \textbf{0.291}\rlap{$^{\ast}$}  &  0.413  &  0.422  &  \textbf{0.984}  &  0.368  &  0.541  &  0.804  & \bf 0.983  &  0.260  &  0.390  &  0.753  &  \textbf{0.986}  & \bf 0.284 & 0.473 & 0.443 & \bf  0.985 \\
\addlinespace[0.5em]
Qwen3  &  NEST  &  0.119  &  0.298  &  0.525  &  0.979  &  0.008  &  0.205  &  - &  0.998  &  0.036  &  0.202  &  0.493  &  0.990  &  0.048  &  0.201  &  0.858  &  0.995  &  0.022  &  0.151  &  0.840  &  0.994  & 0.053 & 0.227 & 0.487 & 0.990 \\
  &  SEMQA  &  \textbf{0.533}  &  \textbf{0.628}  &  \textbf{0.551}  &  0.964  &  0.025  &  \textbf{0.349}  &  - &  0.624  &  0.358  &  \textbf{0.557}  &  \textbf{0.584}  &  0.939  &  0.397  &  \textbf{0.531}  &  \textbf{0.885}  &  0.958  &  \textbf{0.381}  &  \textbf{0.518}  &  \textbf{0.870}  &  0.941  & \bf 0.328 & \bf 0.516 & \bf 0.515 & 0.871 \\
  &  CHyD (ours)  &  0.514  &  0.615  &  0.539  &  \textbf{1.000}  &  \textbf{0.026}  &  0.287  &  -  &  \textbf{1.000}  &  \textbf{0.362}\rlap{$^{\ast}$}  &  0.546  &  0.558  &  \textbf{0.998}  &  \textbf{0.400}\rlap{$^{\ast}$}  &  0.521  &  \textbf{0.885}  &  \textbf{1.000}  &  0.380  &  0.490  &  0.865  &  \textbf{1.000}  & 0.326 & 0.492 & 0.506 & \bf 1.000 \\
\bottomrule
\end{tabular}
}
\end{table*}

\subsection{Implementation Details}

All methods are evaluated using three state-of-the-art models with few-shot prompting:
\begin{itemize}
    \item \texttt{Qwen3-4B-Instruct-2507} (referred to as Qwen3)\footnote{\url{https://huggingface.co/Qwen/Qwen3-4B-Instruct-2507}} \cite{qwen3technicalreport};
    \item \texttt{Llama-3.1-8B-Instruct} (referred to as Llama-3.1)\footnote{\url{https://huggingface.co/meta-llama/Llama-3.1-8B-Instruct}} \cite{grattafiori2024llama};
    \item \texttt{Mistral-Nemo-Instruct-2407} (referred to as Mistral-N)\footnote{\url{https://huggingface.co/mistralai/Mistral-Nemo-Instruct-2407}} \cite{mistral2024nemo}.
\end{itemize}
This selection aims to demonstrate that CHyD is model-agnostic and consistent across different architectures. 
Prompts instruct models to answer based on the provided context and to output \textit{``No answer''} when the context does not contain sufficient information to answer the query, preventing forced extraction when no faithful span exists in the source documents. For SEMQA and CHyD, models are instructed to place extracted evidence between brackets. By contrast, NEST is prompted to generate answers only based on the given context, without explicit citation formatting. For the purely extractive baseline, we also include the encoder-only DistilBERT\footnote{\url{https://huggingface.co/distilbert/distilbert-base-cased-distilled-squad}}.

To distinguish the generation behaviour from the extraction correctness across the various approaches, we provide only the gold context for each example, without additional retrieval, across the baseline datasets (MedMCQA, MESAQA and QuoteSum). This isolates the decoding mechanisms from retrieval-induced errors.

However, in safety-critical domains, Question-Answering systems must remain robust against irrelevant information. Therefore, for our technical dataset, SEMAeroSQuAD, we evaluate two distinct settings to test this robustness:
\begin{itemize}
    \item \textbf{Single-context}: The model is provided with the gold context exclusively.
    \item \textbf{Multi-context}: The model is provided with the gold context alongside 4 distractor passages, which are retrieved using the \texttt{all-MiniLM-L6-v2} embedding model\footnote{\url{https://huggingface.co/sentence-transformers/all-MiniLM-L6-v2}}.
\end{itemize}
This distinction is necessary to demonstrate that CHyD maintains its extraction faithfulness even when the LLM faces retrieval noise, a critical requirement in real-world applications.

CHyD introduces an Extraction of Evidence from the Context (EEC) mode, triggered when the model generates the activation token \texttt{``[''}. At inference time, we construct a suffix tree of the tokenised retrieved contexts, which contains all suffixes of the given text. While in EEC mode, the model may either continue copying valid context suffixes based on the already generated span or generate a special token to exit EEC mode and resume free generation. To account for the tokeniser's specific behaviour and allow more fluent answers, we treat tokens such as \texttt{``]''}, \texttt{``].''} and \texttt{``],''} as valid for exiting EEC mode.

Generation parameters were selected to balance answer diversity and fluency. The maximum number of tokens to generate is set based on the average answer length for each dataset. Additionally, we used 1-2 few-shot examples per dataset to provide minimal in-context guidance on the expected answer format. Table \ref{tab:generation_config} reports the generation parameters used for each dataset, including Temperature (Temp.), Nucleus sampling threshold (Top-p), the maximum number of tokens to generate in the answer (Max Tokens), and the number of few-shot examples (Few-shots). All methods use the same configuration per dataset to ensure fair comparisons. Because the generation process is not deterministic under these parameters, all experiments for the generative models are executed across two independent runs. The results reported in the evaluation tables represent the average of these runs.

\section{Results}

We evaluate our CHyD approach across various QA settings, comparing it against NEST and SEMQA. Our analysis focuses on three complementary aspects:
\begin{enumerate}
    \item Extraction correctness and faithfulness;
    \item Answer quality and fluency;
    \item Fluency-faithfulness trade-offs.
\end{enumerate}

To provide a comprehensive summary of model behaviour across diverse scenarios, Tables \ref{tab:extractive_eval} and \ref{tab:fluency_eval} include an ``Overall'' column. This column reports the average of each metric computed across the four primary evaluation settings (MedMCQA, MESAQA, QuoteSum and the single-context configuration of SEMAeroSQuAD). The multi-context setting of SEMAeroSQuAD is excluded from this general average to prevent double-weighting this dataset and to maintain a balanced representation of the main configurations. The detailed results show that while generation-oriented baselines may achieve higher fluency scores, they fail to guarantee faithful evidence extraction, a requirement that our approach satisfies by construction.

\begin{table*}[ht]
\caption{Answer quality and fluency across the datasets. Bold numbers indicate the best performance. $\dagger$ and $\ast$ denote significant improvement of CHyD over SEMQA and NEST, respectively ($p < 0.05$). RL: ROUGE-L, SQA: SEMQA Score, M.: MAUVE, BF1: BERTScore-F1, Lat.: Latency (in seconds). Note: Llama-3.1 refers to Llama-3.1-8B-Instruct, Mistral-N refers to Mistral-Nemo-Instruct-2407, and Qwen3 refers to Qwen3-4B-Instruct.}
\label{tab:fluency_eval}
\centering
\resizebox{\textwidth}{!}{%
\begin{tabular}{ll | rrrrr | rrrrr | rrrrr | rrrrr | rrrrr | rrrrr}
\toprule
\bf LLM & \bf Approach & \bf RL & \bf SQA & \bf M. & \bf BF1 & \bf L. (s) & \bf RL & \bf SQA & \bf M. & \bf BF1 & \bf L. (s) & \bf RL & \bf SQA & \bf M. & \bf BF1 & \bf L. (s) & \bf RL & \bf SQA & \bf M. & \bf BF1 & \bf L. (s) & \bf RL & \bf SQA & \bf M. & \bf BF1 & \bf L. (s) & \bf RL & \bf SQA & \bf M. & \bf BF1 & \bf L. (s) \\
\midrule
\rowcolor{gray!10} & & \multicolumn{5}{c|}{\textbf{MedMCQA}} & \multicolumn{5}{c|}{\textbf{MESAQA}} & \multicolumn{5}{c|}{\textbf{QuoteSum}} &  \multicolumn{10}{c|}{\textbf{SEMAeroSQuAD}} & \multicolumn{5}{c}{\textbf{Overall}} \\
\rowcolor{gray!10} & & \multicolumn{5}{c|}{} & \multicolumn{5}{c|}{} & \multicolumn{5}{c|}{} &  \multicolumn{5}{c|}{single-context} &  \multicolumn{5}{c|}{multi-context} &  \multicolumn{5}{c}{} \\
\addlinespace[0.5em]
DistilBERT  &   &  0.355  &  0.356  &  0.184  &   0.613  &  0.002  &  0.089  &  0.122  &  0.069  &   0.532  & 0.009  & 0.213  &  0.353  &  0.065  &   0.543  & 0.007  &   0.294  &  0.478  &  0.068  &  0.581  & 0.004  &  0.256  &  0.403  &  0.123   &  0.564  & 0.009 & 0.238 & 0.327 & 0.097 & 0.567 & 0.005 \\
\addlinespace[0.5em]
Llama-3.1  &  NEST  &  0.107  &  0.151  &  0.023  &  0.479  &  1.825  &  \textbf{0.445}  &  0.328  &  \textbf{0.771}  &  \textbf{0.728}  &  2.441  &  0.531  &  0.280  &  0.955  &  0.740  &  5.949  &  0.600  &  \textbf{0.333}  &  0.847  &  0.791  &  1.423  &  0.604  &  \textbf{0.285}  &  0.880  &  0.802  &  1.384 & 0.421 & 0.273 & 0.649 & 0.684 & 2.909 \\
  &  SEMQA  &  0.160  &  0.236  &  0.048  &  0.516  &  0.889  &  0.430  &  \textbf{0.337}  &  0.763  &  0.720  &  \textbf{1.592}  &  \textbf{0.537}  &  \textbf{0.420}  &  0.950  &  \textbf{0.748}  &  \textbf{2.111}  &  0.662  &  0.250  &  \textbf{0.962}  &  \textbf{0.833}  &  \textbf{0.594}  &  \textbf{0.620}  &  0.225  &  0.926  &  \textbf{0.809}  &  \textbf{0.759} & \bf 0.447 & \bf 0.311 & \bf 0.681 & \bf 0.704 & \bf 1.296 \\
  &  CHyD (ours)  &  \textbf{0.172}\rlap{$^{\ast\dagger}$}  &  \textbf{0.256}\rlap{$^{\ast\dagger}$}  &  \textbf{0.057}  &  \textbf{0.528}\rlap{$^{\ast\dagger}$}  &  \textbf{0.814}  &  0.424  &  0.296  &  0.750  &  0.713  &  6.014  &  0.522  &  0.402  &  \textbf{0.959}  &  0.737  &  3.748  &  \textbf{0.663}\rlap{$^{\ast}$}  &  0.258  &  0.948  &  0.831  &  0.776  &  0.618  &  0.232  &  \textbf{0.931}  &  0.807  &  4.246 & 0.445 & 0.303 & 0.678 & 0.702 & 2.838 \\
\addlinespace[0.5em]
Mistral-N  &  NEST  &  0.200  &  0.222  &  0.117  &  0.550  &  2.831  &  \textbf{0.441}  &  0.341  &  0.573  &  0.723  &  3.724  &  \textbf{0.479}  &  0.273  &  0.855  &  0.714  &  9.485  &  \textbf{0.515}  &  0.363  &  \textbf{0.573}  &  \textbf{0.728}  &  2.348  &  \textbf{0.508}  &  0.305  &  \textbf{0.588}  &  \textbf{0.736}  &  2.203 & 0.409 & 0.300 & 0.529 & 0.679 & 4.597 \\
  &  SEMQA  &  0.352  &  0.373  &  0.474  &  0.658  &  0.536  &  0.431  &  \textbf{0.405}  &  \textbf{0.850}  &  \textbf{0.730}  &  \textbf{2.470}  &  0.469  &  \textbf{0.369}  &  0.854  &  \textbf{0.726}  &  \textbf{2.757}  &  0.490  &  \textbf{0.461}  &  0.535  &  0.718  &  \textbf{0.654}  &  0.496  &  \textbf{0.392}  &  0.584  &  0.728  &  \textbf{0.861} & 0.435 & \bf 0.402 & 0.678 & 0.708 & \bf 1.604 \\
  &  CHyD (ours)  &  \textbf{0.453}\rlap{$^{\ast\dagger}$}  &  \textbf{0.475}\rlap{$^{\ast\dagger}$}  &  \textbf{0.884}  &  \textbf{0.743}\rlap{$^{\ast\dagger}$}  &  \textbf{0.466}  &  0.397  &  0.359  &  0.792  &  0.709  &  6.583  &  0.459  &  0.360  &  \textbf{0.889}  &  0.713  &  4.216  &  0.465  &  0.402  &  0.497  &  0.706  &  0.941  &  0.461  &  0.291  &  0.565  &  0.711  &  4.418 & \bf 0.444 & 0.399 & \bf 0.766 & \bf 0.718 & 3.051 \\
\addlinespace[0.5em]
Qwen3  &  NEST  &  \textbf{0.299}  &  0.246  &  \textbf{0.213}  &  \textbf{0.599}  &  2.595  &  \textbf{0.427}  &  0.282  &  0.462  &  0.695  &  3.487  &  0.527  &  0.294  &  0.885  &  0.720  &  4.978  &  0.609  &  0.311  &  0.613  &  0.780  &  1.453  &  0.600  &  0.261  &  0.588  &  0.782  &  1.412 & 0.466 & 0.283 & 0.543 & 0.699 & 3.128 \\
  &  SEMQA  &  0.228  &  \textbf{0.362}  &  0.089  &  0.570  &  \textbf{0.575}  &  0.420  &  \textbf{0.313}  &  \textbf{0.564}  &  \textbf{0.706}  &  \textbf{2.132}  &  \textbf{0.556}  &  \textbf{0.510}  &  \textbf{0.896}  &  \textbf{0.752}  &  \textbf{2.056}  &  \textbf{0.683}  &  \textbf{0.469}  &  \textbf{0.891}  &  \textbf{0.833}  &  \textbf{0.614}  &  \textbf{0.664}  &  \textbf{0.466}  &  \textbf{0.862}  &  \textbf{0.821}  &  \textbf{0.765} & \bf 0.472 & \bf 0.413 & \bf 0.610 & \bf 0.715 & \bf 1.344 \\
  &  CHyD (ours)  &  0.221  &  0.352  &  0.090  &  0.569  &  0.601  &  0.408  &  0.267  &  0.542  &  0.699  &  6.414  &  0.537  &  0.492  &  0.840  &  0.738  &  3.866  &  0.678  &  0.459  &  0.885  &  0.831  &  0.812  &  0.655  &  0.437  &  0.807  &  0.818  &  4.323 & 0.461 & 0.393 & 0.589 & 0.709 & 2.923 \\
\bottomrule
\end{tabular}
}
\end{table*}

\subsection{Extraction Correctness and Faithfulness}

Table \ref{tab:extractive_eval} presents extraction-specific results, including Exact Match (EM), F1 score, Answer-Level Recall (ALR), and the proposed Extraction Faithfulness Accuracy (EFA) metric. These metrics capture complementary aspects of extraction behaviour. While EM and F1 score assess answer correctness, EFA measures whether quoted spans are exact continuous substrings of the retrieved context.

CHyD consistently achieves an EFA between 0.981 and 1.000 across all models and datasets, indicating that nearly all quoted spans exactly match continuous spans in the retrieved context. This result confirms that enforcing hard extraction during decoding provides, by construction, a strong and reliable guarantee of evidence faithfulness. The remaining failures appear to stem from rare tokenisation edge cases, where the extraction trigger token is not emitted as an isolated token (e.g. generating the token \texttt{``[S''} instead of the trigger \texttt{``[''}), preventing a switch to the EEC mode. In such cases, the model remains in free generation and generates the quoted span in the same probabilistic manner as SEMQA.

In contrast, SEMQA frequently generates quoted spans that do not appear verbatim in the retrieved context. While it produces fluent, semantically plausible answers on open-domain tasks like QuoteSum, its EFA drops sharply on domain-specific datasets, falling to 0.371 on MESAQA and 0.628 on MedMCQA with Mistral-N. This shows that quoted-generation approaches can create an illusion of evidence grounding while failing to ensure faithful extraction. In addition, our strict extraction approach does not degrade extractive performance relative to SEMQA, demonstrating that faithfulness guarantees can be achieved without sacrificing answer correctness. In fact, on datasets like MedMCQA, enforcing these constraints yields statistically significant improvements ($p < 0.05$) in both EM and F1 for CHyD over SEMQA and NEST when using Llama-3.1 and Mistral-N. With Mistral-N, CHyD achieves an EM of 0.415, strongly outperforming SEMQA (0.309) and NEST (0.136). 

Conversely, NEST's relaxed speculative decoding mechanism enables it to extract exact spans from the source, sometimes yielding a high EFA. However, these extractions are opportunistic and aimed at accelerating inference rather than evidence grounding. Consequently, they do not generally correspond to the supporting evidence, resulting in consistently lower EM and F1 scores than CHyD. Therefore, even though both leverage ``copy-paste'' mechanisms, their different objectives explain NEST's consistently lower EM and F1 scores than CHyD across almost all configurations. 

Low Exact Match scores across all models on MESAQA occur because the EM metric requires the extracted span to match the gold answer exactly. Since gold answers in this abstractive dataset are complete free-form sentences, achieving an exact verbatim match is highly improbable. In such cases, the evaluated approaches tend to quote pertinent keywords rather than entire sentences, which also explains why Answer-Level Recall is not reported for this dataset.

To further contextualise our results, we include the purely extractive DistilBERT baseline. As shown in Table \ref{tab:extractive_eval}, this model achieves near-perfect Extraction Faithfulness Accuracy (0.994 on average) across all datasets, acting as an extractive upper bound. The slight deviation from a perfect 1.000 EFA occurs because DistilBERT predicts extraction boundaries at a subword-token level. When converting these predicted tokens back to text, the tokeniser's decoding often normalises spacing and punctuation, meaning that the decoded string may not always exactly match the source content.

Consequently, it achieves high Exact Match and F1 scores on extraction-friendly datasets like SEMAeroSQuAD. However, its performance collapses on the medical datasets, with Exact Match dropping to 0.022 on MedMCQA and 0.012 on MESAQA, highlighting a critical limitation of purely extractive models. Identifying the correct evidence on complex tasks, such as MedMCQA, requires advanced domain-specific reasoning capabilities that a standard extractive model lacks. Moreover, the Overall averages reveal that our approach, CHyD, combined with modern generalist LLMs, such as Qwen3, can achieve an overall EM score that surpasses the purely extractive DistilBERT performance (0.328 and 0.314, respectively). This demonstrates that CHyD effectively leverages LLMs' superior reasoning capabilities to extract precise evidence.

Overall, these results prove that speculative RAG approaches can be redesigned to enforce strict grounding constraints, providing verifiable faithfulness to the context in safety-critical domains.

\subsection{Answer Quality and Fluency}
 
Table \ref{tab:fluency_eval} reports the overall fluency performance of the evaluated approaches across all datasets, using metrics such as ROUGE-L, BERTScore-F1, and the SEMQA score. 

SEMQA and NEST frequently achieve the highest scores on these generation-oriented metrics across datasets such as MESAQA, QuoteSum and SEMAeroSQuAD. This reflects their architectural optimisation for unconstrained, fluent answer generation and summarisation. Because standard QA metrics reward plausible paraphrasing, these baseline models score highly even when their extracted evidence is technically unfaithful.

Furthermore, the extractive DistilBERT baseline results illustrate the trade-offs of bypassing language generation. While DistilBERT achieves near-zero inference latency (under 0.01 seconds across all configurations), this speed comes at a cost of answer fluency. Because this model can only output unformatted verbatim text spans from the context, it shows drops in distribution-based and generation-oriented metrics. For instance, on QuoteSum and the single-context SEMAeroSQuAD, its MAUVE scores collapse to 0.065 and 0.068, respectively, dropping well below those of the LLM-based approaches (which frequently exceed 0.85). Moreover, its ROUGE-L and BERTScore-F1 remain strictly inferior to those of the generative approaches on abstractive and semi-extractive tasks. This confirms that while pure extractive approaches excel at identifying evidence, they fail to generate grammatically coherent and properly formatted answers required by QA systems in critical domains. 

Although CHyD is not primarily designed to optimise paraphrasing fluency, it remains competitive on standard QA metrics despite enforcing strict extraction constraints during decoding. Furthermore, it improves exact extraction correctness, as discussed previously. On SEMAeroSQuAD (single-context), CHyD achieves a ROUGE-L of 0.663 with Llama-3.1, slightly outperforming SEMQA (0.662). Across almost all configurations, CHyD's BERTScore-F1 remains within a few percentage points of the best-performing baseline, indicating that enforcing faithful evidence reproduction does not lead to a significant drop in answer quality or readability. In fact, the Overall column in Table \ref{tab:fluency_eval} shows that the fluency trade-off is practically non-existent. For instance, with Llama-3.1, the overall ROUGE-L difference between the unconstrained SEMQA and our strictly constrained CHyD is 0.002, whereas with Mistral-N, CHyD achieves higher overall MAUVE and BERTScore-F1 scores.

Moreover, enforcing verbatim extraction via a Suffix Tree increases latency. In tasks with longer contexts (MESAQA and SEMAeroSQuAD), CHyD shows a higher latency than SEMQA. For instance, on MESAQA, CHyD's latency ranges from 6.014 to 6.583 seconds, compared to SEMQA's 1.592 to 2.470 seconds. In safety-critical domains, such as aircraft maintenance, this increased latency is a necessary and highly acceptable trade-off to guarantee procedural compliance. Notably, CHyD still consistently achieves lower latency than NEST on datasets such as QuoteSum.

Overall, these results highlight a significant trade-off in hybrid generation, where optimising fluency does not necessarily imply strong grounding or faithfulness to the retrieved context. In safety-critical domains, where any deviation from procedures is intolerable, faithful decoding and extractive-oriented evaluation criteria are therefore more appropriate than fluency-based metrics alone.

\subsection{Fluency-Faithfulness Trade-offs}

Based on the results in Tables \ref{tab:extractive_eval} and \ref{tab:fluency_eval}, the differences in extraction behaviour stem from the distinct objectives of the evaluated methods. SEMQA is designed to prioritise fluent and plausible answers in a semi-extractive format. Consequently, this approach often achieves high scores on generation-oriented metrics (such as ROUGE-L and MAUVE) even when its quoted spans are not exact verbatim copies of the original documents. These hallucinated quotes remain undetected by standard QA metrics but are effectively captured by the Extraction Faithfulness Accuracy (EFA), exposing a critical vulnerability in ``fluency-first'' approaches.

In contrast, NEST primarily aims to accelerate inference by using a relaxed speculative decoding mechanism that opportunistically copies spans from the context. While this mechanism may sometimes yield exact evidence extraction, it does not guarantee the correctness of the copied evidence. As a result, NEST produces fluent and occasionally faster answers, but with inconsistent grounding guarantees, as reflected in its lower EM and F1 scores.

CHyD adopts a different objective: rather than optimising fluency or latency, we prioritise extraction faithfulness by enforcing hard constraints during decoding whenever evidence is quoted. As a result, CHyD achieves near-perfect Extraction Faithfulness Accuracy (0.981 - 1.000) across all LLMs and datasets. This significantly outperforms SEMQA, whose EFA drops below 0.371 on the technical dataset MESAQA, and NEST, which lacks grounding guarantees. While our strict design may slightly limit the model's stylistic freedom, it ensures that quoted evidence is always extracted from its context. Moreover, the overall results indicate that SEMQA's average EFA is highly model-dependent (ranging from 0.639 with Mistral-N to 0.871 with Qwen3), whereas CHyD maintains a near-perfect, stable EFA (0.985 to 1.000) across the used LLMs.

While the purely extractive baseline, DistilBERT, achieves the highest F1 and EM scores on QuoteSum and SEMAeroSQuAD, this encoder-only model is structurally restricted to outputting raw text spans from the context. As a result, it cannot generate conversational, grammatically correct answers, synthesise complex queries, extract multiple disconnected spans across different passages (as required by the dataset QuoteSum), or naturally integrate evidence into fluent sentences. This limitation is highlighted by its performance drop across all generation-oriented metrics, such as MAUVE and BERTScore-F1. The contribution of CHyD is not to surpass purely extractive models at their tasks, but rather to enable hybrid generation, which is necessary for safety-critical applications. CHyD achieves this by delivering near-perfect extraction faithfulness without sacrificing the global answer fluency.

These results highlight a clear trade-off in hybrid generation. Optimising fluency or inference speed does not necessarily align with strong guarantees of faithfulness. In safety-critical domains like aircraft maintenance or medical QA, where strict compliance with certified documentation is mandatory, prioritising correctness over fluency is a necessary design choice, and a few seconds of latency is an acceptable trade-off for guaranteed procedural compliance.

\subsection{Qualitative Analysis}

Table \ref{tab:examples} presents two representative failures of existing hybrid QA approaches. In Example 1, NEST extracts only the final digit of the numerical answer, producing a fluent but factually incorrect output. In Example 2, SEMQA generates a plausible quoted span (\textit{``anyone in the general public''}) that is absent from the context. Moreover, in this example, NEST correctly answers the query, but the supporting evidence (\textit{``anyone''}) is not among the extracted spans.

In both cases, CHyD produces correct answers while ensuring that quoted spans are exact copies from the source documents. These examples highlight that neither answer fluency nor quoted generation alone can guarantee context faithfulness. Overall, the results show that enforcing faithfulness during decoding changes the behaviour of hybrid QA systems. While existing methods optimise efficiency or fluency, CHyD demonstrates that it is possible to achieve hybrid generation with near-perfect extraction faithfulness, a mandatory requirement in safety-critical applications.

\begin{table}[ht]
    \caption{Qualitative comparison of hybrid QA outputs using the Qwen3 model. Example 1 highlights a numerical extraction error by NEST, while Example 2 shows an unfaithful quoted generation by SEMQA. Verbatim extracted spans of evidence are highlighted in \ctext{blue!40}{blue}, verbatim extracted spans that do not match the evidence in \ctext{blue!15}{light blue}, and generated quotes not present in the context in \ctext{red!30}{red}.}
    \label{tab:examples}
    \centering
    \resizebox{\linewidth}{!}{%
    \begin{tabular}{p{0.1\textwidth}p{0.8\textwidth}}
   \toprule
    \rowcolor{gray!10} \multicolumn{2}{l}{\textbf{Example 1: Numerical extraction error}} \\
    \addlinespace[0.5em]
    Query &
    How many planes were allowed to be operated commercially although not in compliance with FAA safety regulations? \\ \addlinespace[0.5em]
    Context &
    ... Jim Oberstar, former chairman of the committee said its investigation uncovered a pattern of regulatory abuse and widespread regulatory lapses, allowing \ctext{blue!40}{117 aircraft} \ctext{blue!15}{to} be \ctext{blue!15}{operated commercially although not in compliance with FAA safety} rules\ctext{blue!15}{.} Oberstar said ... \\ \addlinespace[0.5em]
    Gold answer &    117 \\ \addlinespace[0.5em]
    NEST & 1\ctext{blue!15}{[1]} aircraft were allowed\ctext{blue!15}{[ to]} be\ctext{blue!15}{[ operated commercially although not in compliance with FAA safety]} regulations\ctext{blue!15}{[.]} \\ \addlinespace[0.5em]
    SEMQA & \ctext{blue!40}{[ 117 aircraft ]} were allowed to be operated commercially although not in compliance with FAA safety rules. \\ \addlinespace[0.5em]
    CHyD (Ours) & \ctext{blue!40}{[117]} planes were allowed to be operated commercially although not in compliance with FAA safety regulations. \\ 
    \midrule
    \rowcolor{gray!10} \multicolumn{2}{l}{\textbf{Example 2: Hallucinated quoted span}} \\
    \addlinespace[0.5em]
    Query & Who could be an air traffic controller irrespective of experience now because of the rule change? \\ \addlinespace[0.5em]
    Context & In 2014, the FAA changed a long-standing approach to air traffic control candidates that eliminated preferences based on training and experience at flight schools in favor of a personality test open to \ctext{blue!40}{anyone} irrespective \ctext{blue!15}{of experience}. The move was made to increase flight \ctext{blue!15}{traffic controller} racial diversity. Before the \ctext{blue!15}{change}, candidates who had completed coursework at participating colleges and universities could \ctext{blue!15}{be} ``fast-tracked'' for consideration... \\ \addlinespace[0.5em]
    Gold answer & anyone \\ \addlinespace[0.5em]
    NEST & Anyone could\ctext{blue!15}{[ be]} an air\ctext{blue!15}{[ traffic controller]} irrespective\ctext{blue!15}{[ of experience]} now because\ctext{blue!15}{[ of]} the rule\ctext{blue!15}{[ change]}. \\ \addlinespace[0.5em]
    SEMQA & Now, \ctext{red!30}{[ anyone in the general public ]} could be an air traffic controller irrespective of experience because of the rule change. \\ \addlinespace[0.5em]
    CHyD (Ours) & Now, \ctext{blue!40}{[ anyone ]} could be an air traffic controller irrespective of experience because of the rule change. \\
    \bottomrule
    \end{tabular}
}
\end{table}

\subsection{Answerability Detection}

Answerability detection is evaluated only on the SEMAeroSQuAD dataset, the only dataset in our experimental setup that contains unanswerable queries. While answerability is not a primary focus of this work, models are explicitly instructed to output \textit{``No answer''} when the provided context lacks sufficient evidence, ensuring that extraction constraints are only applied when faithful spans exist. 

Table \ref{tab:squadaero_answerable} reports the abstention performance of the evaluated approaches across both single-context and multi-context retrieval scenarios. We observe comparable accuracy across all LLMs in detecting unanswerable queries. Notably, Llama-3.1, Mistral-N and Qwen3 maintain strong abstention capabilities, achieving F1 scores above 0.88 across all approaches.

Across nearly all configurations, CHyD achieves answerability detection metrics that are nearly identical to the unconstrained baseline SEMQA. For example, in the multi-context scenario, CHyD and SEMQA achieve an exact tie in accuracy (0.832) and F1 score (0.896) when using Llama-3.1. Similarly, with Qwen3, CHyD maintains highly competitive accuracy across both single- and multi-context scenarios, trailing SEMQA by less than 0.004.

This indicates that enforcing faithfulness constraints via CHyD does not reduce the models' ability to abstain when evidence is missing, even when applied across different model architectures.

\begin{table}[ht]
    \caption{Answerability detection performance on SEMAeroSQuAD (answerable vs unanswerable queries). Prec.: Precision, Rec.: Recall, Acc.: Accuracy. Note: Llama-3.1 refers to Llama-3.1-8B-Instruct, Mistral-N refers to Mistral-Nemo-Instruct-2407, and Qwen3 refers to Qwen3-4B-Instruct.}
    \label{tab:squadaero_answerable}
    \centering
    \resizebox{\columnwidth}{!}{%
    \begin{tabular}{ll |rrrr|rrrr}
    \toprule
        \bf Dataset / LLM & \bf Approach & \bf Prec. & \bf Rec. & \bf Acc. & \bf F1 &   \bf Prec. & \bf Rec. & \bf Acc. & \bf F1 \\
    \midrule
        \rowcolor{gray!10} \multicolumn{2}{l|}{\textbf{SEMAeroSQuAD}} & \multicolumn{4}{c|}{single-context} &  \multicolumn{4}{c}{multi-context} \\
        \addlinespace[0.5em]
        Llama-3.1 & NEST & 0.850 & 0.973 & 0.852 & 0.908 & 0.800 & \bf 0.989 & 0.806 & 0.885\\
                  & SEMQA & \bf 0.880 & \bf 0.975 & \bf 0.878 & \bf 0.923 & \bf 0.840 & 0.963 & \bf 0.832 & \bf 0.896 \\
                  & CHyD (ours) & 0.870 & 0.965 & 0.867 & 0.916 & \bf 0.840 & 0.964 & \bf 0.832 & \bf 0.896  \\ \addlinespace[0.5em]
        Mistral-N & NEST & 0.820 & 0.990 & 0.832 & 0.899 & 0.790 & \bf 0.996 & 0.796 & 0.880 \\
                  & SEMQA & \bf 0.840 & 0.993 & \bf 0.849 & \bf 0.908 & \bf 0.800 & 0.992 & \bf 0.811 & \bf 0.888 \\
                  & CHyD (ours) & 0.810 & \bf 0.994 & 0.824 & 0.895 & 0.790 & 0.995 & 0.801 & 0.883  \\ \addlinespace[0.5em]
        Qwen3 & NEST & 0.790 & \bf 0.993 & 0.800 & 0.882 & 0.820 & \bf 0.991 & 0.828 & 0.897  \\
              & SEMQA & \bf 0.870 & 0.985 & \bf 0.880 & \bf 0.925 & \bf 0.860 & 0.979 & \bf 0.862 & \bf 0.914  \\
              & CHyD (ours) & \bf 0.870 & 0.984 & 0.876 & 0.923 & \bf 0.860 & 0.977 & 0.861 & \bf 0.914  \\ 
    \bottomrule
    \end{tabular}
    }
\end{table}

\section{Discussion and Limitations}

This work explicitly prioritises faithfulness over fluency or latency, reflecting mandatory requirements in safety-critical environments, like aircraft maintenance, where answers must strictly adhere to certified documentation. In such settings, properties such as traceability, compliance, and exact reproduction of evidence are mandatory, not optional. Any deviation from the source documentation, even if the output is fluent, can have severe operational consequences. 

Enforcing strict decoding constraints inevitably limits the model's expressive freedom and may reduce fluency or abstraction. In particular, constraining generation to verbatim extraction when creating quoted spans reduces paraphrasing flexibility and may lead to less fluent answers. However, our results show that this constraint eliminates a significant source of hallucination while preserving competitive answer quality. In safety-critical domains, correctness must take precedence over stylistic considerations, and our results demonstrate that this trade-off is necessary.

CHyD's first limitation is its reliance on the switching conditions. While extraction is guaranteed to be faithful once triggered, the model must still learn when to enter EEC mode. In rare cases, the model may skip or delay the extraction trigger (\texttt{``[''}) or produce token variants, leading to unquoted or unextracted spans. Future work could mitigate this limitation by fine-tuning the model with an explicit objective for mode switching and introducing dedicated control tokens into the vocabulary (e.g. \texttt{``<EEC>''} and \texttt{``</EEC>''}) to improve robustness and answer fluency. A second limitation concerns the model's reduced ability to paraphrase or summarise evidence due to the hard extraction constraints, potentially making CHyD less suited for open-domain settings or creative applications. 

Additionally, strict decoding constraints increase CHyD's latency compared with SEMQA. While our approach is not optimised for inference time, this increase is due to the evidence extraction phases and scales with the number of quoted spans rather than the length of the answer. This suggests that faithfulness guarantees can be achieved without sacrificing the benefits of hybrid generation. Finally, while SEMAeroSQuAD enables controlled evaluation of faithfulness in closed-domain settings, it is automatically generated and may therefore introduce biases from the generation process. Consequently, complementary human evaluation would be valuable for assessing answer fluency, comprehensiveness, and correctness.

Finally, a critical distinction must be made between extractive faithfulness and reasoning faithfulness. CHyD provides a hard guarantee for the former: any content within brackets is a verbatim, verified span from the context. However, the reasoning (i.e. the decision of which span to extract) still resides with the base LLM. While CHyD eliminates hallucination within the evidence itself, the logical link between the query and the quoted span remains a subject for future research.

Overall, our work demonstrates that speculative RAG, previously explored primarily for inference speed-up, can be reoriented to enforce strict evidence extraction. While approaches such as NEST focus on probabilistic span acceptance to improve efficiency, our results suggest that combining speculative RAG with strict faithfulness constraints could offer a promising direction for future research. In particular, strategies that integrate fast speculative decoding with constrained extraction may help improve efficiency while preserving correctness guarantees.

\section{Conclusion}
Generative AI has proven to be a highly effective approach for technical question answering. However, in safety-critical use cases where faithfulness guarantees are required, standard generative paradigms fall short. In this work, we showed that existing hybrid and semi-extractive QA approaches can produce fluent answers while altering the quoted spans, creating an illusion of evidence grounding without guaranteeing strong faithfulness to the retrieved context. In particular, methods such as SEMQA may modify quoted spans to improve fluency, citing, in some cases, spans that do not appear in the source documents. Such behaviour is unacceptable in safety-critical applications like aircraft maintenance, where answers must strictly match the certified documentation.

We presented Constrained Hybrid Decoding (CHyD), a framework that enforces verbatim extraction at decoding time. By shifting the objective of hybrid generation from fluency or efficiency to correctness guarantees via strong extraction constraints, we provide a principled solution for safety-critical QA. While this framework yields a slight trade-off in fluency and modest reductions in generation-oriented metrics, such as ROUGE-L and BERTScore, it provides strong guarantees on evidence faithfulness.

Our results demonstrate that hybrid generation can be made suitable for safety-critical domains by prioritising evidence correctness over fluency and latency. Although CHyD is not optimised for inference speed, we aim, in future work, to combine existing speculative decoding methods with faithfulness constraints to further improve efficiency. Overall, this work presents a concrete hybrid generation framework that meets the strict requirements of high-stakes domains, such as aircraft maintenance, thereby enabling the reliable deployment of Large Language Models in applications where faithfulness is non-negotiable.

\bibliographystyle{ACM-Reference-Format}
\bibliography{_biblio}


\end{document}